\documentclass[preprint,amsmath,amssymb,nofootinbib]{revtex4}
\usepackage[normalem]{ulem}
\usepackage{changes}
\usepackage{amsfonts}
\usepackage{color}
\usepackage[colorlinks=true,
linkcolor=blue,
breaklinks=true,
urlcolor=blue,
citecolor=blue]{hyperref}

\usepackage{amsmath,graphicx,color,epsfig,slashed}

\newcommand{\nn}{\nonumber}
\newcommand{\be}{\begin{equation}}
\newcommand{\ee}{\end{equation}}
\newcommand{\bea}{\begin{eqnarray}}
\newcommand{\eea}{\end{eqnarray}}
\newcommand{\ba}{\begin{array}}
\newcommand{\ea}{\end{array}}
\newcommand{\bi}{\begin{itemize}}
\newcommand{\ei}{\end{itemize}}

\newcommand{\mcb}{{\mathcal B}}

\newcommand{\nslash}{\kern 0.2 em n\kern -0.50em /}
\newcommand{\kslash}{\kern 0.2 em k\kern -0.45em /}
\newcommand{\qslash}{\kern 0.2 em q\kern -0.45em /}
\newcommand{\pslash}{\kern 0.2 em p\kern -0.50em /}
\newcommand{\rslash}{\kern 0.2 em r\kern -0.50em /}
\newcommand{\sslash}{\kern 0.2 em s\kern -0.50em /}
\newcommand{\Sslash}{\kern 0.2 em S\kern -0.50em /}
\newcommand{\Pslash}{\kern 0.2 em P\kern -0.50em /}
\newcommand{\Dslash}{\kern 0.2 em D\kern -0.65em /\kern 0.15em}

\newcommand{\uestc}{\affiliation{School of Physics, University of Electronic Science and
Technology of China, Chengdu 610054, China}}

\newcommand{\ucas}{\affiliation{University of Chinese Academy of Sciences, Beijing 100049, China}}

\newcommand{\imp}{\affiliation{Institute of Modern Physics, Chinese Academy of Sciences, Lanzhou 730000, China}}

\begin{document}
\title{Hunting for the heavy quark spin symmetry partner of $Z_{cs}$} 

\author{Xu Cao}\email{caoxu@impcas.ac.cn}
\imp
\ucas

\author{Zhi Yang}\email{zhiyang@uestc.edu.cn (corresponding author)}
\uestc

\begin{abstract}
  The discovery of a charged strange hidden-charm state $Z_{cs}$(3985) implies another higher $Z^*_{cs}$ state coupling to $\bar{D}_s^{*-}D^{*0} + c.c$ under the heavy quark spin symmetry.
  In this paper we discuss a possible hunt for it with data taken at existing facilities.
  We point out a hint for $Z^*_{cs}$ in the data of $\bar{B}_s^0 \to J/\psi  K^- K^+$ at LHCb, though weak, in line with the production mechanism of pentaquark $P_c$.
  We also study the triangular singularity which would possibly enhance the production of $Z^*_{cs}$ in electron-positron collision. 
  Surprisingly, the production rate of $Z^*_{cs}$ is expected to be maximum at the $e^+ e^-$ center of mass energy of 4.648 GeV, which is lower than 4.681 GeV for $Z_{cs}$ due to the inverted coupling hierarchy of $D_{s1} \bar D K$ and $D_{s2} \bar D^* K$ in the triangle diagrams.
  Their bottom analogue under heavy quark flavor symmetry is also discussed. 
  Our theoretical analysis would confront with future experiment of LHCb, BESIII, and Bell II.
\end{abstract}
\maketitle

\section{Introduction} \label{sec:intro}

The evidence of strange pentaquark $P_{cs}$(4459) from LHCb Collaboration \cite{LHCb:2020jpq}, though only 3$\sigma$ significance, marked the dawn of strange era for the exploration of exotic candidates. Soon afterwards the BESIII collaboration discovered a charged resonance with the significance of more than 5$\sigma$ in the $D_s^-D^{*0} + D_s^{*-}D^{0}$ mass distribution of $e^+ e^- \to K^+ (D_s^-D^{*0} + D_s^{*-}D^{0})$, whose mass and width are~\cite{BESIII:2020qkh}
\bea \label{eq:dataBES} \nn
Z_{cs}(3985)&:& \quad 3982.5^{+1.8}_{-2.6} \pm {2.1} \,\textrm{MeV}, \quad 12.8^{+5.3}_{-4.4} \pm {3.0} \,\textrm{MeV} 
\eea
It is close to the $D_s^-D^{*0}$  and $D_s^{*-}D^{0}$ thresholds with the probable quantum number being $J^P = 1^+$.
This motivates its interpretation of $D_s^-D^{*0} + D_s^{*-}D^{0}$ molecular state as the strange partner of $Z_c$(3900) in various scenarios, e.g. in QCD sum rules before~\cite{Lee:2008uy,Dias:2013qga} and after its observation~\cite{Wang:2020htx,Ozdem:2021hka,Ozdem:2021yvo,Wang:2020iqt,Azizi:2020zyq,Wang:2020dgr,Xu:2020evn,Wan:2020oxt}. 
Its inner structure and mass spectra are extensively investigated in hadro-quarkonium~\cite{Voloshin:2019ilw} and compact tetraquark models~\cite{Ferretti:2020ewe}, and various quark models, for instance, the chiral quark model~\cite{Yang:2021zhe,Chen:2021uou}, 
the dynamical diquark model \cite{Giron:2021sla,Shi:2021jyr},
the model with meson-meson and diquark-antidiquark constituents \cite{Jin:2020yjn,Ebert:2008kb}.
Its  decays of hidden charm \cite{Wu:2021ezz} and open charm~\cite{Chen:2013wca} channels, configuration mixing \cite{Karliner:2021qok} and compositeness \cite{Guo:2020vmu} are also studied.

Whether one-Boson-exchange potential is strong enough to bind $D_s^-D^{*0}$  and $D_s^{*-}D^{0}$ is inconclusive in the coupled channel formalism after considering the OZI suppression \cite{Sun:2020hjw,Chen:2020yvq,Ding:2021igr}.
Other binding mechanisms for molecular explanation are under investigation, e.g. contact interaction \cite{Yang:2020nrt,Meng:2020ihj,Ikeno:2020mra,Dong:2020hxe,Dong:2021bvy,Dong:2021juy,Du:2020vwb}, 
axial-meson exchange \cite{Yan:2021tcp}, and channel recoupling mechanism \cite{Simonov:2020ozp}.
Under SU(3)-flavor symmetry, its existence is definitely expected as the strange partner of $Z_c$(3900), theorized to be isovector $D^* \bar{D}^*$ molecular. 
Another higher strange axial-vector state $Z_{cs}$(4130) of $D_s^* \bar{D}^*$ component (labeled as $Z_{cs}^*$ hereafter), as strange partner of $Z_c^*$(4020) of $D^* \bar{D}^*$ molecular nature, is predicted in many scenarios~\cite{Yang:2020nrt,Meng:2020ihj,Du:2020vwb,Zhu:2021vtd}.
So a complete multiplets under heavy quark flavor and spin symmetry (HQFS and HQSS) is emerging after the evident twin molecules $Z_{c}$(3900) 
 and $Z_{c}$(4020)
in hidden charm sector~\cite{Nieves:2012tt,HidalgoDuque:2012pq,Guo:2013sya,Guo:2017jvc}, and their close analogs $Z_{b}$(10610) and $Z_{b}$(10650) in hidden bottom sector~\cite{Guo:2013sya}. 

In QCD sum rule~\cite{Dias:2013qga}, initial $K$-meson emission mechanism~\cite{Chen:2013wca}, and hadrocharmonium picture~\cite{Voloshin:2019ilw}, hidden-charm channels are anticipated to be essential for understanding the $Z_{cs}$(3985). 
However, $e^+ e^- \to J/\psi K^- K^+$ is not statistically achievable at present \cite{Belle:2007dwu,Belle:2014fgf,BESIII:2018iop}.
Recent data of $B^+ \to J/\psi \phi K^+$ from the LHCb Collaboration reveals the existence of two wide charged strange hidden-charm states in the $J/\psi K^+$ spectrum~\cite{LHCb:2021uow}. The Breit-Wigner (BW) masses and widths are, respectively
\bea  \label{eq:dataLHC1}  \nn
Z_{cs}(4000)&:& \quad 4003 \pm 6^{+4}_{-14} \,\textrm{MeV}, \quad 131 \pm 15 \pm 26 \,\textrm{MeV}
\\  \label{eq:dataLHC2}  \nn
Z_{cs}(4220)&:& \quad 4216 \pm 24 ^{+43}_{-30} \,\textrm{MeV}, \quad 233 \pm 52^{+97}_{-73} \,\textrm{MeV}
\eea
These $Z_{cs}$ states are unambiguously assigned to be $J^P = 1^+$ states.
The masses of $Z_{cs}$(3985) and $Z_{cs}$(4000) are consistent within uncertainties, motivating the hypothesis of them as one state \cite{Yang:2020nrt,Ortega:2021enc}.
In this case, the $Z_{cs}$(4220) would be an excited state of $Z_{cs}$(3985).
If considering them as two different states, the $Z_{cs}(3985) \to J/\psi K$ decay is suppressed in the limit of the HQSS \cite{Meng:2021rdg}, and the $Z_{cs}$(4220) at LHCb would be the expected axial-vector $D_s^* \bar{D}^*$ molecular. 
Instead, the $Z_{cs}^*$ is predicted to be a $D_s^* \bar{D}^*$ resonance of tensor $2^+$ nature with mass of about 4126 MeV and width of 13 MeV.
In both schemes a narrow $Z_{cs}^*$ close to $D_s^* \bar{D}^*$ threshold plays a central role.
In Sec.~\ref{sec:hints} a weak hint is shown for this state in the data of $\bar{B}_s^0 \to J/\psi  K^- K^+$ at LHCb.
A wide $Z_{cs}$(4250), consistent with $Z_{cs}$(4220) at LHCb, is predicted by HQSS within hadro-quarkonium framework~\cite{Voloshin:2019ilw} together with another $Z_{cs}$(4350) as the respective partners of $Z_c$(4100)~\cite{LHCb:2018oeg} and $Z_c$(4200)~\cite{Belle:2014nuw} (for a discussion of their spin-parity see~\cite{Cao:2018vmv}).

In electron-positron collision, kinematic effects, e.g. triangle singularity (TS)~\cite{Chen:2013wca,Yang:2020nrt} and kinematic reflection \cite{Wang:2020kej} are thought to play an important role. 
When all the intermediate particles in a triangle diagram move collinearly on their mass shell a triangle singularity happens to mimic the resonance-like peak in the invariant mass distributions~\cite{Guo:2019twa}. 
The width of particles in triangle diagrams of $B$-decay would induce the wide $Z_{cs}$ at LHCb~\cite{Ge:2021sdq}, explaining the $Z_{cs}$(3985)/$Z_{cs}$(4220) width difference.
It is possible to distinguish TS mechanism by searching for the photo- \cite{Cao:2020cfx} and lepto-production \cite{Yang:2021jof} and pion/kaon induced reactions \cite{Liu:2021ojf} of exotic mesons, because the on-shell condition is hardly satisfied in these reactions~\cite{Liu:2016dli,Cao:2019kst}.
On the other hand, the triangle diagram possess a triangle singularity under a special energy, hence it becomes an amplifier of the production of exotic candidates in certain energies as experiments found.
The $Z_{c}$(3900) signal is enhanced by the $D_1 \bar{D}^0 D^*$ triangle diagram when the $e^+ e^-$ center of mass (c.m.) energy is around Y(4260), close to the $D_1 \bar{D}^0$ threshold \cite{Albaladejo:2015lob}.
Analogously the $D_{s2}(2573) \bar{D}_s^* D_0$ triangle diagram would enhance the production of $Z_{cs}$(3985) at c.m. energy of 4.681 GeV~\cite{Yang:2020nrt}.
Considering that the $D_{s1}$(2536) is approximately the HQSS partner of the $D_{s2}$(2573),
we point out in Sec.~\ref{sec:Frwk} that the production of predicted axial-vector $Z_{cs}^*$ would be magnified by the $D_{s1}(2536) \bar{D}_s^* D_0^*$ triangle around $D_{s1}(2536) \bar{D}_s^*$ threshold.

\section{A hint of $Z_{cs}^{*}$ at $\bar{B}_s$-decay  } \label{sec:hints}

\begin{figure}
  \begin{center}
  {\includegraphics*[width=0.45\textwidth]{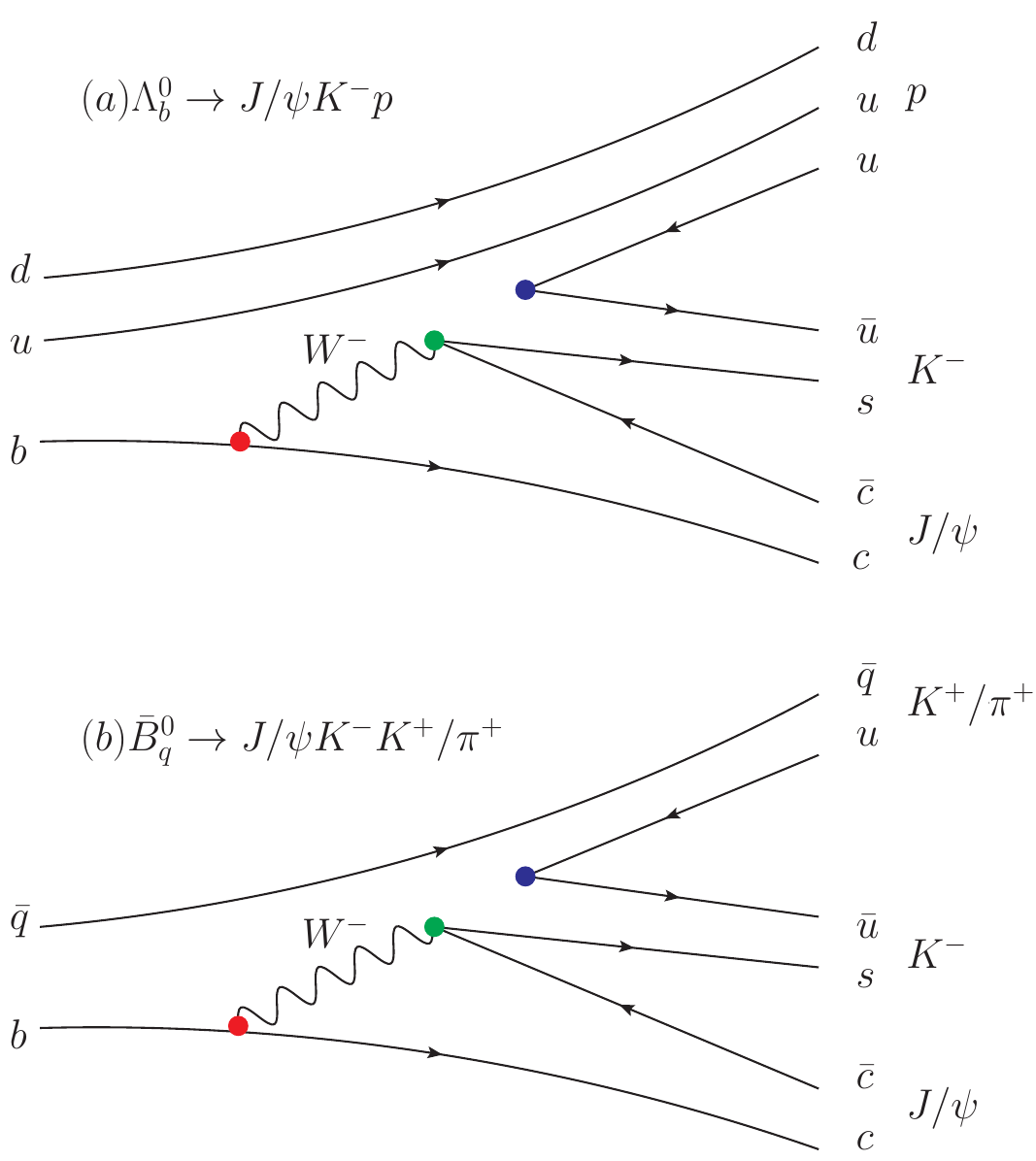}}
    \caption{Internal $W$-emission in (a) the $\Lambda_b \to J/\psi  K^- p$ decay, and (b) the $\bar{B}^0 \to J/\psi  K^- \pi^+$ and $\bar{B}_s^0 \to J/\psi  K^- K^+$. The diagram for the charge conjugate channel of $B^+ \to J/\psi \phi K^+$ can be obtained with $u\bar{u} \to s\bar{s}$ in (b).}
    \label{fig:internal}
  \end{center}
\end{figure}
\begin{figure}
  \begin{center}
  {\includegraphics*[width=0.45\textwidth]{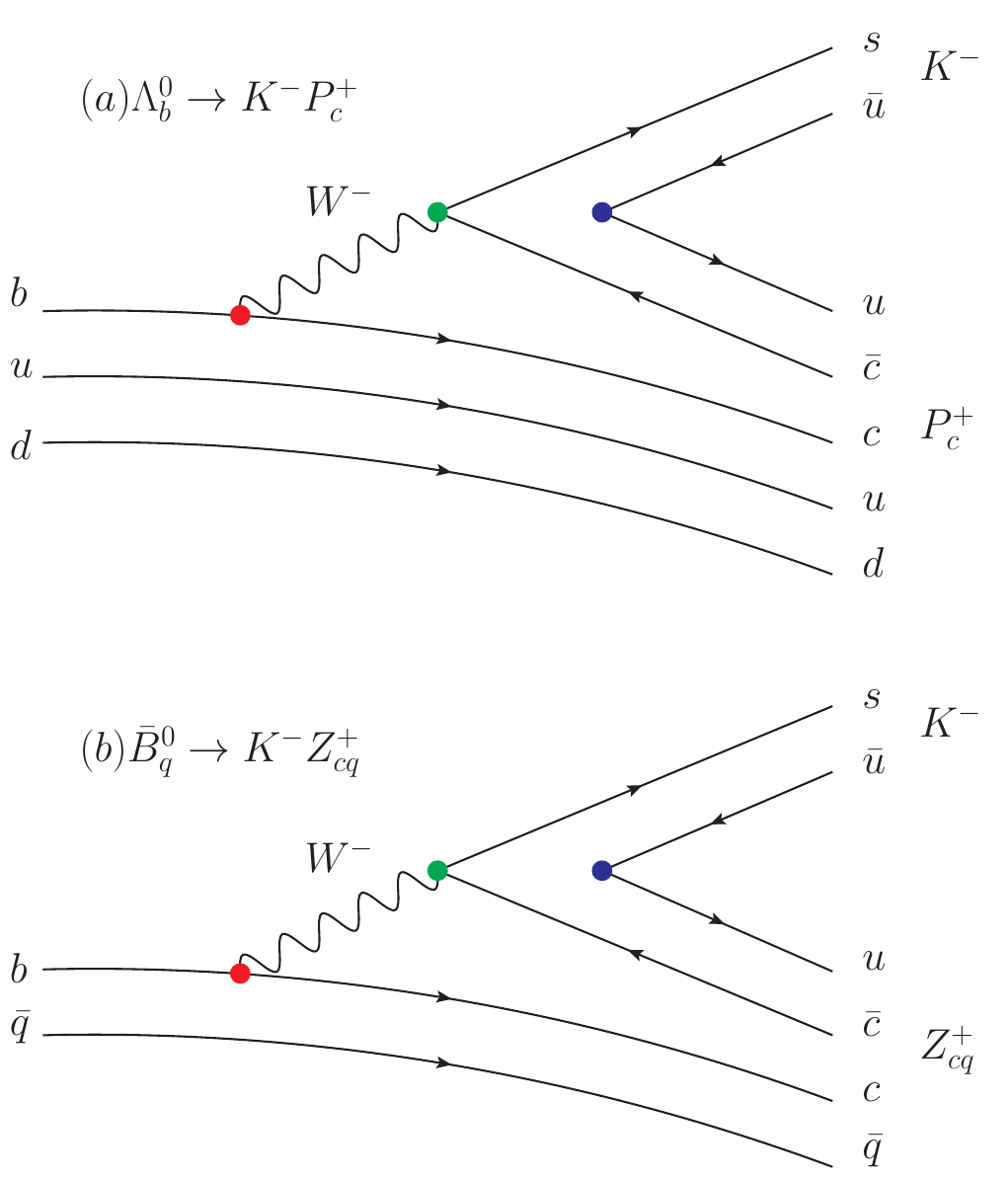}}
    \caption{Exotic states production by external $W$-emission in (a) the $\Lambda_b \to P_c^+ K^-$ decay, and (b) the $\bar{B}^0 \to Z_c^+  K^- \pi^+$ and $\bar{B}_s^0 \to Z_{cs}^{(*)}  K^-$. The diagram for the charge conjugate channel of $B^+ \to Z_{cs}^+ \phi $ can be obtained with $u\bar{u} \to s\bar{s}$ in (b).}
    \label{fig:external}
  \end{center}
\end{figure}
\begin{figure}
  \begin{center}
  {\includegraphics*[width=0.55\textwidth]{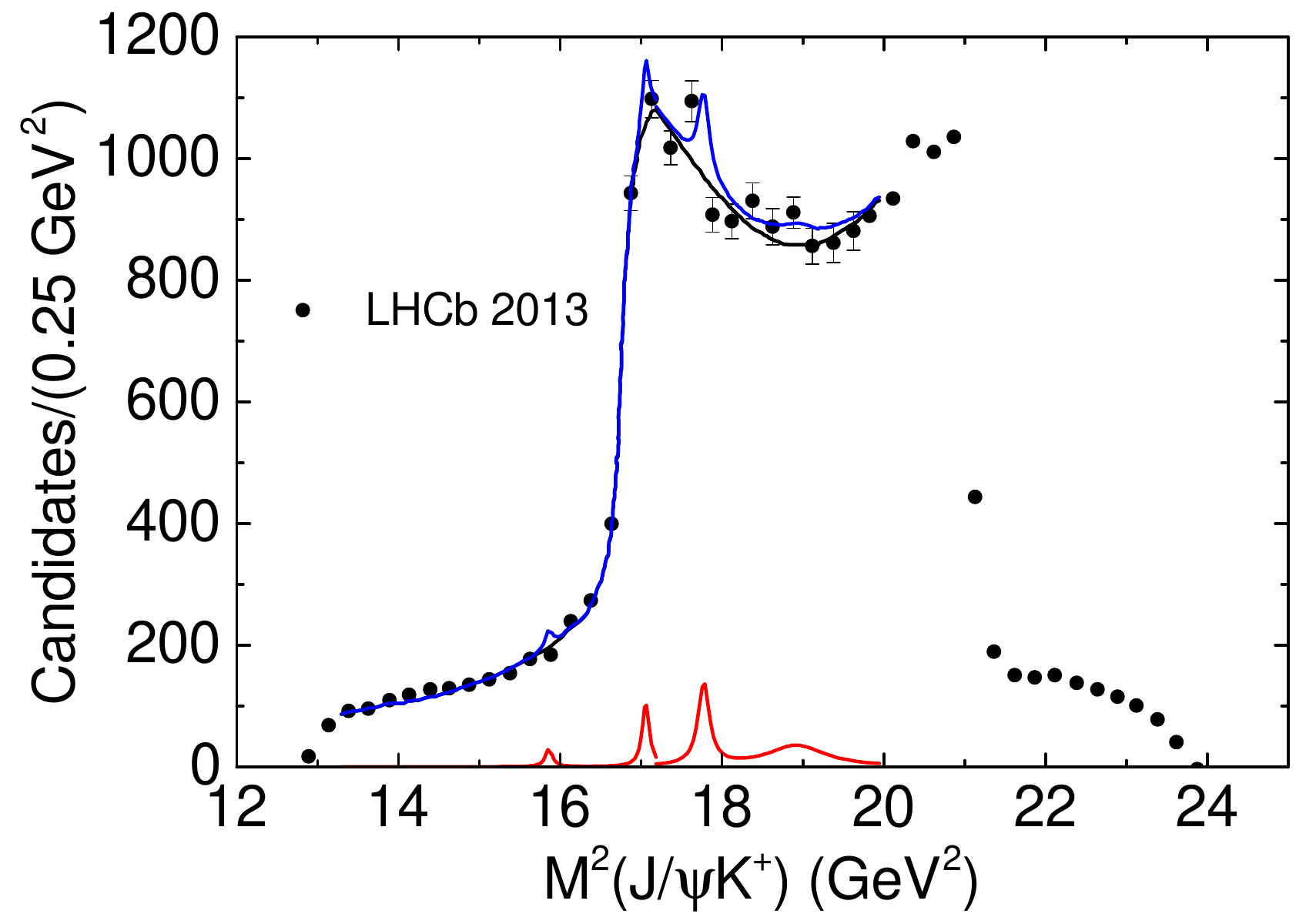}}
    \caption{ The data of $J/\psi K^+$ spectrum in $\bar{B}_s^0 \to J/\psi  K^- K^+$ at LHCb \cite{LHCb:2013kpp} and the possible contribution of $Z_{cs}$(3985), $Z_{cs}^*$(4130), $Z_{cs}$(4220) and $Z_{cs}$(4350). The red curves are Breit-Wigner (BW) distributions of those $Z_{cs}$. Here the central values of mass and width from BESIII are used for $Z_{cs}$(3985). The central values of mass from LHCb is used for $Z_{cs}$(4220) with a width of 20.0 MeV. The mass and width of $Z_{cs}^*$(4130) are respectively 4130 MeV and 15 MeV as expected by HQSS and SU(3)-flavor symmetry~\cite{Wang:2020htx,Meng:2020ihj,Meng:2021rdg}. The mass of $Z_{cs}$(4350) is quoted from hadro-quarkonium~\cite{Voloshin:2019ilw} and width 100 MeV is adopted. The green curve is the incoherent sum of BW and simulated background mainly reflected from $K^- K^+$ spectrum by LHCb.}
    \label{fig:Zcspectrum}
  \end{center}
\end{figure}

The hidden-charm pentaquark states $P_c$ with the same light quark content as the nucleon are discovered in $\Lambda_b \to J/\psi  K^- p$ by LHCb collaboration \cite{LHCb:2015yax,LHCb:2019kea}.
Based on SU(3)-flavor symmetry, the bottom baryon decay would be decomposed into the internal and external $W$-emission diagrams \cite{Cheng:2015cca}.
The three quarks $c\bar{c}s$ produced directly from the $b$-quark decay by the internal $W$-emission in Fig.~\ref{fig:internal}(a) are too energetic to form a bound pentaquark. 
The dominant pentaquark production process is anticipated to be the external $W$-emission amplitudes in Fig.~\ref{fig:external}(a) \cite{Cheng:2015cca}.
The close resemble diagrams for bottom meson decays are shown in Fig.~\ref{fig:internal}(b) and Fig.~\ref{fig:external}(b) for $\bar{B}^0 \to J/\psi  K^- \pi^+$ and $\bar{B}_s^0 \to J/\psi  K^- K^+$.
The $Z_c$(4200) and $Z_c$(4600) are already discovered in the former decay by LHCb \cite{LHCb:2019maw,Belle:2014nuw}, supporting that this argument of production mechanism is similarly applicable for bottom meson decays. 
Another decay of this beneficial feature is $B^+ \to J/\psi \phi K^+$, in which two wide $Z_{cs}$ appeared after considerably increasing statistics at LHCb~\cite{LHCb:2021uow} in comparison with previous measurements \cite{LHCb:2016axx,LHCb:2016nsl,CDF:2009jgo,CDF:2011pep,CMS:2013jru}.
So it seems that $\bar{B}_s^0 \to J/\psi  K^- K^+$ deserve further search for $Z_{cs}$ production though it bears low statistics at present \cite{LHCb:2013kpp}.
Current data of this decay at LHCb in Fig.~\ref{fig:Zcspectrum} shows none clue for $Z_{cs}(3985)/Z_{cs}(4000)$, but does hint weakly for $Z_{cs}^{*}$ and $Z_{cs}(4220)$, both of which are narrow. Another wide 4350 MeV can be accommodated, but its signal is even more fuzzy.
Due to the wide energy bins (500 MeV) of the data, their evidence is rather inconclusive.

In an isospin analysis of $B\to D^*\bar{D}K$ it is shown that the production of the isospin
triplet state $Z_c$(3900) is highly suppressed in $B$ decays compared to the isospin singlet $X$(3872) \cite{Yang:2017nde}. 
As the strange partner of $Z_c$(3900), $Z_{cs}$(3985) would be also absent in the { open-channel} $B$ decays by isospin suppression.
Similar TS as discussed in $B^+ \to J/\psi \phi K^+$ \cite{Ge:2021sdq} would also raise the width of $Z_{cs}$(3985) and $Z_{cs}^{*}$ to be around 100 MeV, making them hard to appear in $B$- and $\bar{B}_s$-decays.
But the masses in the triangles are demanding for TS due to the fixed masses of mother particles $B$/$B_s$. 

Considering the branching ratios $\mcb(\bar{B}^0 \to J/\psi  K^- \pi^+) = (1.15\pm{0.05})\times 10^{-3}$, $\mcb(B^+ \to J/\psi \phi K^+) = (5.0\pm{0.4})\times 10^{-5}$, and $\mcb(\bar{B}_s^0 \to J/\psi  K^- K^+) = (2.54\pm{0.35})\times 10^{-6}$~\cite{Zyla:2020zbs}, increase of the accumulated statistics by around 20 times at least would induce the refined discovery of $Z_{cs}^{*}$ and others.

\section{Triangle diagrams in $e^+ e^-$ annihilation} \label{sec:Frwk}

\begin{table}
  \begin{center}
 \begin{tabular}{c|c|c}
\hline\hline
       & Virtual  & Resonant  \\
\hline
  $C^{(O)}$ (fm$^2$) & $-0.77^{+0.12}_{-0.10}\,\left(-0.45^{+0.05}_{-0.04}\right)$     &   $-0.72^{+0.18}_{-0.13} \, \left(-0.44^{+0.06}_{-0.05}\right)$     \\
  $D^{(O)}$ (fm$^4$) & -   &   $-0.17^{+0.21}_{-0.21}\,
  \left(-0.025^{+0.066}_{-0.049}\right)$     \\
\hline
  $Z_{cs}$ (MeV)    &  $3974_{-3}^{+2}\,\left(3971^{+3}_{-6}\right)$  &       $3963^{+15}_{-5} - i 3^{+17}_{-3}\,\left(3966^{+13}_{-31} - i 0^{+31}_{-0}\right)$\\
  $Z_{cs}^*$ (MeV)  &   $4117_{-5}^{+3}\,\left(4115^{+3}_{-6}\right)$       &   $4110^{+11}_{-5} - i 0^{+15}_{-0}\,\left(4111^{+10}_{-23} - i 0^{+28}_{-0}\right)$     \\
\hline\hline
   \end{tabular}
  \end{center}
  \caption{Parameters of contact interaction and poles of $Z_{cs}^{(*)}$ taken from Ref.~\cite{Yang:2020nrt}. Results outside (inside) brackets are for cutoff $\Lambda=0.5
\;\text{GeV}\;(1\;\text{GeV})$, respectively.
  \label{tab:JPC}}
\end{table}

\begin{figure}
  \begin{center}
  {\includegraphics*[width=0.4\textwidth]{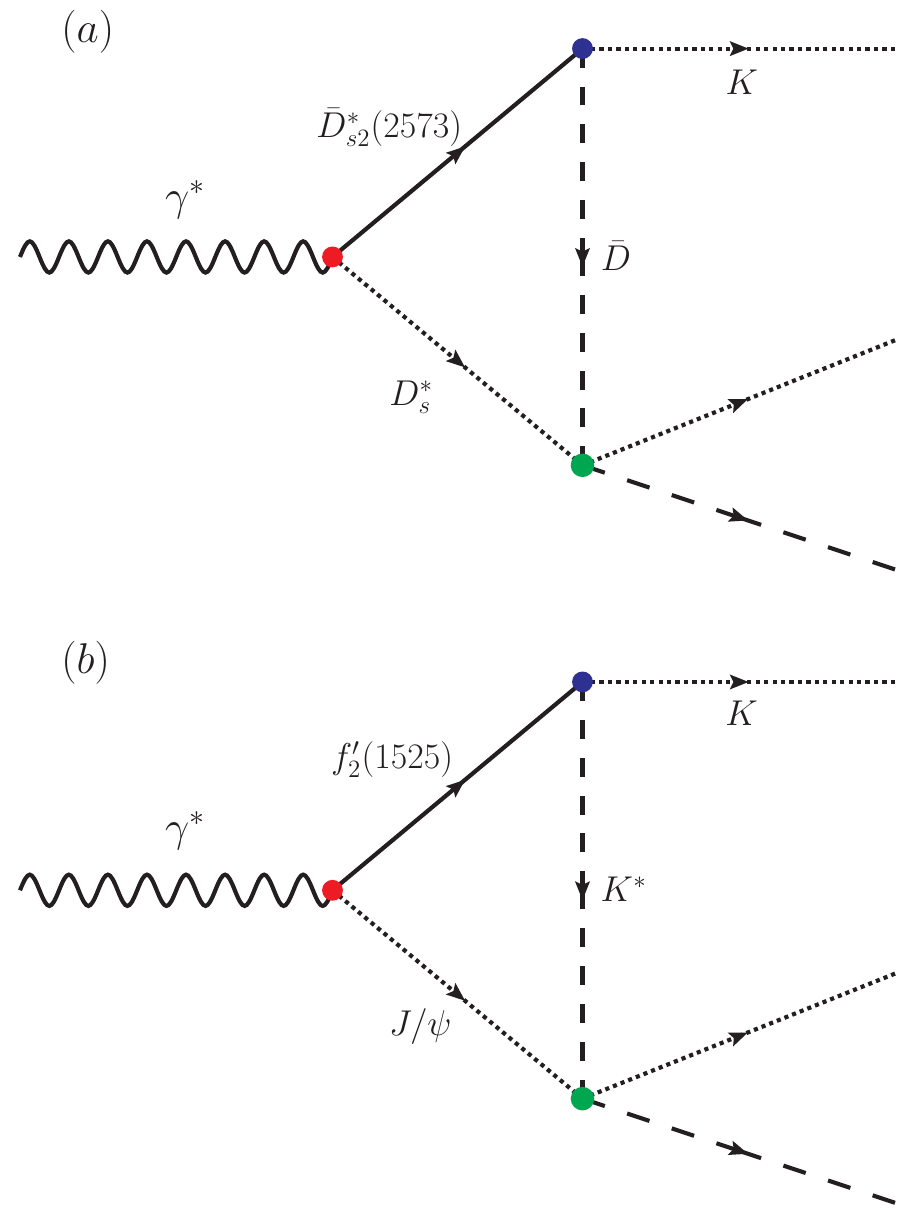}}
    \caption{Possible (a) $D_{s2}(2573) \bar{D}_s^* D_0$ and (b) $J/\psi f_2^{'}(1525) K^*$ triangle diagrams of $e^+ e^- \to K^+ Z_{cs}$(3985) process with the generated $Z_{cs}$(3985) (green circles) decaying to $D_s^-D^{*0}$, $D_s^{*-}D^{0}$, $J/\psi K$ or other possible channels.
    The blue vertices are all in $D$-wave here.
    The green circles denote the $T$-matrix elements which include the effects of the generated $Z_{cs}$(3985) state.} 
    \label{fig:triZcs1}
  \end{center}
\end{figure}

\begin{figure}
  \begin{center}
  {\includegraphics*[width=0.4\textwidth]{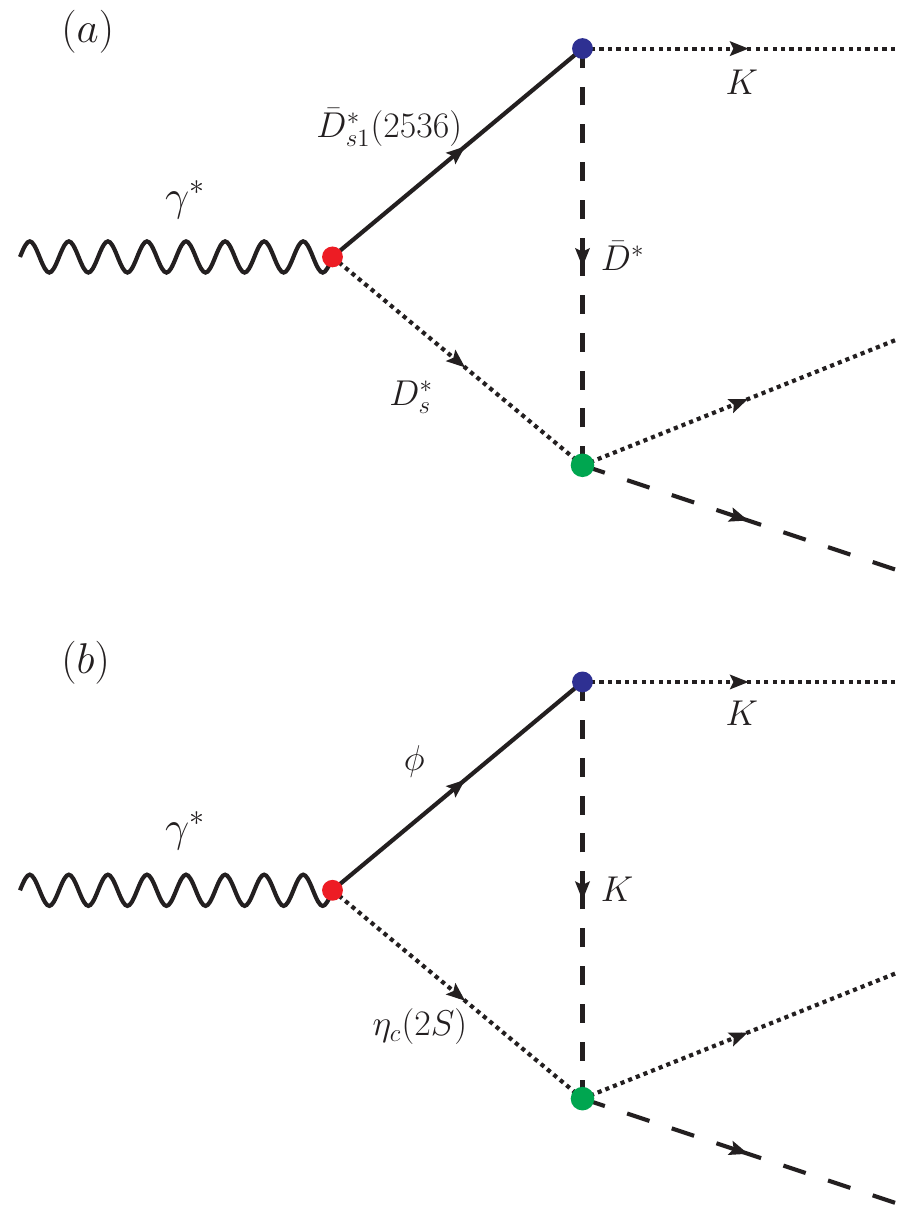}}
    \caption{Possible (a) $D_{s1}(2536) \bar{D}_s^* D_0^*$ and (b) $\eta_c \phi K$ triangle diagrams of $e^+ e^- \to K^+ Z_{cs}^*$ process with the generated $Z_{cs}^*$ (green circles) decaying to $\bar{D}_s^* D_0$, $J/\psi K$ or other possible channels.
    The green circles denote the $T$-matrix elements which include the effects of the generated $Z_{cs}^*$ state.} 
    \label{fig:triZcs2}
  \end{center}
\end{figure}


\begin{figure}[tb]
    \centering
    \includegraphics[width=0.65\textwidth]{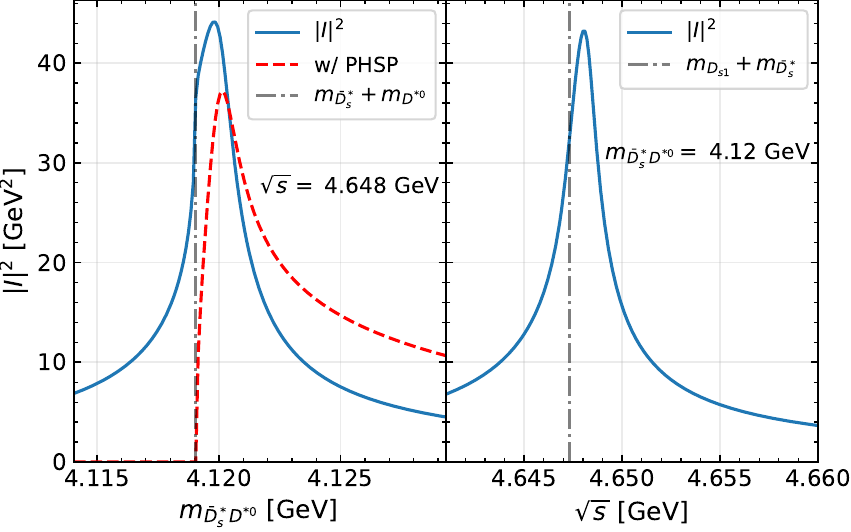}
    \caption{Absolute squared value  of the scalar triangle loop integral, $|I|^2$, with the $D_{s1}\bar D_s^* D^{*0}$ triangle diagram shown as Fig.~\ref{fig:triZcs2}(a). Left: dependence on the $\bar D_s^* D^{*0}$ invariant mass for $\sqrt{s}=4.648$~GeV, where we also show $|I|^2$ convoluted with the phase space, with the maximum normalized to that of $|I|^2$; right: dependence on $\sqrt{s}$ with $m_{\bar D_s^* D^{*0}}= 4.12$~GeV.}
    \label{fig:zsstar}
\end{figure}

The BEPCII continues to run for collecting integrated luminosity at different energies so BESIII provides the chance to look for $Z_{cs}^*$ in $e^+ e^-$ annihilation.
In an effective field theory, the production of $Z_c$(3900/4020) and $Z_{cs}$(3985) are dynamically understandable in a consistent manner with the help of Lippmann-Schwinger equation --- $T = V + V G_0 T$,
{ where $V$ denotes the potential and $G_0$ is the two-point loop function.
For explicit expressions, we refer to Ref.~\cite{Albaladejo:2015lob,Yang:2020nrt} within the nonrelativistic approximation. 
The contact-range interaction incorporating the SU(3)-flavor symmetry contains at most two low-energy constants (LECs) at leading order \cite{Yang:2020nrt},}
\bea
  V^{(O)}_{\rm virtual} &=& C^{(O)}(\Lambda) \, . \label{eq:pot-virtual} \\
  V^{(O)}_{\rm res} &=& C^{(O)}(\Lambda) + 2 D^{(O)}(\Lambda)\,k^2\, , \label{eq:pot-res}
\eea
with $k$ the c.m. momentum of the two mesons.
The momentum independent kernel in Eq.~(\ref{eq:pot-virtual}) generates a bound or a virtual pole below its respective two-meson threshold, while the latter in Eq.~(\ref{eq:pot-res}) generates a resonant state.
Here $\Lambda$ means that {the divergent loop function $G_0$} has been regulated by a Gaussian form factor $e^{-(p/\Lambda)^2}$ with $p$ the momentum of the two mesons in $e^+ e^-$ system. 
The values of LECs can be determined by the masses and widths of $Z_{c}$(3900/4020) and $Z_{cs}$(3985), as summarized in Table \ref{tab:JPC} together with the generated poles of $Z_{cs}^{(*)}$. 

In this framework, the $D_1 \bar{D}^0 D^*$ triangle diagram, as part of the generation of 
the $Z_{c}$(3900) signal, enhances its production when the $e^+ e^-$ center of mass (c.m.) energy is around $Y$(4260), close to the $D_1 \bar{D}^0$ threshold \cite{Albaladejo:2015lob}.
Analogously the $D_{s2}(2573) \bar{D}_s^* D_0$ triangle diagram, contributing to the generation of $Z_{cs}$(3985) as shown in Fig. \ref{fig:triZcs1}(a), would enhance the production of $Z_{cs}$(3985) at the c.m. energy of 4.681 GeV~\cite{Yang:2020nrt}. 
Note that $D_{s2} D \bar{K}$ decay proceeds in $D$-wave. Considering that the $D_{s1}$(2536) is approximately the HQSS partner of the $D_{s2}$(2573), Fig. \ref{fig:triZcs2}(a) with $D_{s1}(2536) \bar{D}_s^* D_0^*$ triangle is a close analog of Fig. \ref{fig:triZcs1}(a). 
The scalar 3-point loop integral of these diagrams is given by~\cite{Guo:2010ak,Guo:2017jvc,Guo:2020oqk}
\begin{widetext}
\begin{align}
  I = \frac{\mu_{12}\mu_{23}}{2\pi\sqrt{a}} \left[
\arctan\left(\frac{c_2-c_1}{2\sqrt{a(c_1-i\epsilon)}}\right) 
- \arctan\left(\frac{c_2-c_1-2a}{2\sqrt{a(c_2-a-i\epsilon)}}\right)
\right],
    \label{eq:Iexp}
\end{align}
\end{widetext}
where $\mu_{12}$ and $\mu_{23}$ are the reduced masses of the $D_{s1}\bar D^*_s$ and $\bar D_s^*D^{*0}$, respectively,
$a = \left(\mu_{23}q_K/m_{D^{*0}}\right)^2$, $c_1= 2\mu_{12}b_{12}$,
$c_2=2\mu_{23}b_{23}+q_K^2\mu_{23}/m_{D^{*0}}$ with $b_{12} = m_{D_{s1}}+m_{\bar D_s^*}-\sqrt{s}$ and 
$b_{23}=m_{\bar D_s^*}+m_{D^{*0}}+E_K-\sqrt{s}$, and $q_K (E_K)$ is the $K^+$ momentum (energy) in the $\gamma^*$ c.m. frame. 
The involved kinematic variables are given by
\begin{align}
    q_K & = \frac1{2M}\sqrt{\lambda(s,m_K^2, m_{23}^2)}, \notag\\
    m_{13}^2 &= m_K^2 + m_{D^{*0}}^2 + 2 E_1^* E_3^* - 2 p_1^* p_3^* \cos\theta_3^*, \notag\\
    p_1^* &= \sqrt{E_1^{*2} - m_K^2},\quad p_3^* = \frac1{2m_{23}}\sqrt{\lambda(m_{23}^2, m_2^2, m_3^2)}, \notag\\
    E_3^* & = \frac{m_{23}^2 - m_2^2 + m_3^2}{2m_{23}}, \quad E_1^* = \frac{s - m_{23}^2 - m_K^2}{2m_{23}}, 
    \quad
\end{align}
where $m_2 = m_{D_{s}^*}$ and $m_3 = m_{D^{*0}}$.

Fig. \ref{fig:zsstar} shows the absolute value squared of the corresponding scalar triangle loop integral $|I|^2$ at the $e^+ e^-$ c.m. energy of 4.648 GeV, around the $D_{s1}\bar{D}_s^*$ threshold.
Quite close results are obtained for different input of parameters from Table \ref{tab:JPC}.
Data are needed to constrain the relative normalization of the different amplitudes, so at present the $m_{\bar D_s^* D^{*0}}$ spectrum can not be predicted.
However, after convolution with the three-body phase space of $e^+ e^- \to \bar D_s^* D^{*0} K^+$ a clear peak around 4.12 GeV appears to enhance the production of axial vector $Z_{cs}^*$. 
The $Z_{cs}^*$ of tensor nature is suppressed in this mechanism due to its involvement of $D$-wave at least for the total production amplitudes, so it serves a possible experimental criteria of $Z_{cs}^*$ with $J^P$. 
Interestingly $\gamma^* D_{s1}\bar{D}_s^*$ vertex in $D$-wave results into 33.0 MeV lower c.m. energy of $Z^*_{cs}$ production than that of the $Z_{cs}$(3985). 
This is due to the inverted coupling hierarchy of $D_{s1} \bar D K$ and $D_{s2} \bar D^* K$.
Unfortunately BEPCII scans the c.m. energies with 20 MeV interval, and its best luminosities are at 4.640 and 4.660 GeV, {a little shift from the best energy to identify the contribution of triangle diagrams. The Bell experiment collects a large data sample at or near the $\Upsilon$ resonances in $e^+e^-$ collision, with the possibility to search for the $Z_{cs}^{(*)}$ state through initial-state radiation~\cite{Belle-II:2018jsg}. }

Other triangle diagrams are possibly present in the production of $Z_{cs}^{(*)}$ in $e^+ e^-$ annihilation.
For the $Z_{cs}$(3985), a triangle of $J/\psi f_2^{'}(1525) K^*$ in Fig. \ref{fig:triZcs1}(b) has singularity in c.m. energy of around 4622 MeV. 
Due to the 86 $\pm$ 5 MeV width of $f_2^{'}(1525)$ and its moderate coupling of $K^* \bar K$, the enhancement would not so sharp and is possibly submerged by the smooth background.
No enhancement of $Z_{cs}$(3985) production in the $e^+ e^- \to K^+ (D_s^-D^{*0} + D_s^{*-}D^{0})$ data of BESIII at 4628 MeV seems to support these arguments~\cite{BESIII:2020qkh}.
Similarly the $\eta_c(2S) \phi K$ triangle in Fig. \ref{fig:triZcs2}(b) would contribute to the $Z_{cs}^*$ production with all vertices in $p$-wave. The narrow width of $\eta_c(2S)$ and strong coupling of $\phi$ to $K \bar K$ would enhance the $Z_{cs}^*$ production at c.m. energy of 4657 MeV, about 10 MeV above that of Fig. \ref{fig:triZcs2}(a).
However, the $Z_{cs}^{(*)}$ in the molecular scenario mainly decay to open charm channels $\bar D_s^{(*)}D^{*0} + D_s^{*} \bar D^{(*)}$, so the $T$-matrix elements denoted by green circles in Figs. \ref{fig:triZcs1} and \ref{fig:triZcs2} generate $Z_{cs}^{(*)}$ state in the elastic channel.
Both Figs. \ref{fig:triZcs1}(b) and \ref{fig:triZcs2}(b) are disadvantage of the inelastic interaction of hidden charm $J/\psi K^*$ or $\eta_c(2S) K$ vertices in $e^+ e^- \to K^+ (\bar D_s^{(*)}D^{*0} + D_s^{*} \bar D^{(*)})$. 

As a result, $D_{s2}(2573) \bar{D}_s^* D_0$ and $D_{s1}(2536) \bar{D}_s^* D_0^*$ triangle diagrams are anticipated to be the dominant singularity for the $Z_{cs}$(3985) and $Z_{cs}^{*}$ production, respectively. Their correspondence in bottom sector is quite similar. The $Z_b(10610)$ and $Z_b(10650)$ ($Z_b$ and $Z_b^*$) discovered a decade ago by the Belle collaboration, are a pair of charged hidden-bottom resonances with $I^G(J^{PC}) = 1^-(1^{+-})$ and masses~\cite{Belle:2011aa}
\bea
  M(Z_b^{\pm}) &=& 10607.2 \pm 2.0 \, {\rm MeV} \, , \\
  M(Z_b^{*\pm}) &=& 10652.2 \pm 1.5 \, {\rm MeV} \, ,
\eea
very close to the $B\bar{B}^*$ and $B^* \bar{B}^*$ thresholds, respectively. Their strange partners are predicted to be close to the $B_s\bar{B}^* + B_s^*\bar{B}$ and $B_s^* \bar{B}^*$ thresholds~\cite{Yang:2020nrt,Cao:2020cfx,Meng:2020ihj,Wang:2020htx}:
\bea
  M(Z_{bs}^{\pm}) &\simeq& 10700 \, {\rm MeV} \, , \\
  M(Z_{bs}^{*\pm}) &\simeq& 10745 \, {\rm MeV} \, ,
\eea
if the $Z_b$'s were indeed bound states of the bottom mesons mentioned above.
Considering that the dominant decays are $B_{s1}(5830) \to B^* K$ and $B_{s2}(5840) \to B K$,
the $B_{s2}(5840) \bar{B}_s^* B$ and $B_{s1}(5830) \bar{B}_s^* B^*$ triangle diagrams are anticipated to be the dominant singularity for the $Z_{bs}$ and $Z_{bs}^{*}$ production in $e^+ e^-$ annihilation, respectively. 
The inverted c.m. energy gap for their production is only about 10 MeV due to the mass difference between $B_{s1}(5830)$ and $B_{s2}(5840)$. 



\section{Summary and Conclusion} \label{sec:summ}

The $Z_{cs}(3985)$ state, together with the $X_0$(2900) and $X_1$(2900) of quark content $\bar{c} \bar{s} u d$ by LHCb~\cite{Aaij:2020hon,Aaij:2020ypa}, raised the question of existence of a complete exotic spectrum of the charm and charm-strange. 
A missing piece of this molecule jigsaw puzzle is $Z_{cs}^{*}$ as expected by HQSS and SU(3)-flavor symmetry.
Inspired by the dominance of external $W$-emission, in this paper we call attention to a hint for $Z^*_{cs}$ production by the same mechanism in the data of $\bar{B}_s^0 \to J/\psi  K^- K^+$ at LHCb, which needs further investigation with higher statistics.
We further explore the triangular singularity which would possibly enhance the production of axial-vector $Z_{cs}^*$ state in $e^+ e^- \to K^{*+} D_s^{*-}D^{*0}$. 
The $Z_{cs}^*$ is expected to be enhanced by the $D_{s1}(2536) \bar{D}_s^* D_0^*$ triangle diagrams, similar to that of $D_{s2}(2573) \bar{D}_s^* D_0$ for $Z_{cs}(3985)$ production $e^+ e^-$ annihilation.
As a surprising result, $Z^*_{cs}$ would be produced in about 30 MeV lower c.m. energy of $e^+ e^-$ than that of $Z_{cs}$, due to the inverted coupling hierarchy of $D_{s1} \bar D K$ and $D_{s2} \bar D^* K$.
The same scenario is considered within alternative interaction kernel~\cite{Baru:2021ddn}.
Their bottom partners are analogous with a relative smaller energy gap of $e^+ e^-$ between $Z_{bs}$ and $Z^*_{bs}$ production. 
Our results are helpful for the future hunt for $Z_{cs}^{*}$ in $\bar{B}_s$-decay and $e^+ e^-$ annihilation.





\begin{acknowledgments}

We would like to thank Juan Nieves, Manuel Pavon Valderrama, Feng-Kun Guo and Jian-Ping Dai for useful communication.
This work is supported by the National Natural Science Foundation of China (Grants Nos. 12075289 and U2032109) and the Strategic Priority Research Program of Chinese Academy of Sciences (Grant NO. XDB34030301).

\end{acknowledgments}

\bibliography{zcs.bib}

\begin{thebibliography}{78}
\expandafter\ifx\csname natexlab\endcsname\relax\def\natexlab#1{#1}\fi
\expandafter\ifx\csname bibnamefont\endcsname\relax
  \def\bibnamefont#1{#1}\fi
\expandafter\ifx\csname bibfnamefont\endcsname\relax
  \def\bibfnamefont#1{#1}\fi
\expandafter\ifx\csname citenamefont\endcsname\relax
  \def\citenamefont#1{#1}\fi
\expandafter\ifx\csname url\endcsname\relax
  \def\url#1{\texttt{#1}}\fi
\expandafter\ifx\csname urlprefix\endcsname\relax\def\urlprefix{URL }\fi
\providecommand{\bibinfo}[2]{#2}
\providecommand{\eprint}[2][]{\url{#2}}

\bibitem[{\citenamefont{Aaij et~al.}(2021{\natexlab{a}})}]{LHCb:2020jpq}
\bibinfo{author}{\bibfnamefont{R.}~\bibnamefont{Aaij}} \bibnamefont{et~al.}
  (\bibinfo{collaboration}{LHCb}), \bibinfo{journal}{Sci. Bull.}
  \textbf{\bibinfo{volume}{66}}, \bibinfo{pages}{1391}
  (\bibinfo{year}{2021}{\natexlab{a}}), \eprint{2012.10380}.

\bibitem[{\citenamefont{Ablikim et~al.}(2021)}]{BESIII:2020qkh}
\bibinfo{author}{\bibfnamefont{M.}~\bibnamefont{Ablikim}} \bibnamefont{et~al.}
  (\bibinfo{collaboration}{BESIII}), \bibinfo{journal}{Phys. Rev. Lett.}
  \textbf{\bibinfo{volume}{126}}, \bibinfo{pages}{102001}
  (\bibinfo{year}{2021}), \eprint{2011.07855}.

\bibitem[{\citenamefont{Lee et~al.}(2009)\citenamefont{Lee, Nielsen, and
  Wiedner}}]{Lee:2008uy}
\bibinfo{author}{\bibfnamefont{S.~H.} \bibnamefont{Lee}},
  \bibinfo{author}{\bibfnamefont{M.}~\bibnamefont{Nielsen}}, \bibnamefont{and}
  \bibinfo{author}{\bibfnamefont{U.}~\bibnamefont{Wiedner}},
  \bibinfo{journal}{J. Korean Phys. Soc.} \textbf{\bibinfo{volume}{55}},
  \bibinfo{pages}{424} (\bibinfo{year}{2009}), \eprint{0803.1168}.

\bibitem[{\citenamefont{Dias et~al.}(2013)\citenamefont{Dias, Liu, and
  Nielsen}}]{Dias:2013qga}
\bibinfo{author}{\bibfnamefont{J.~M.} \bibnamefont{Dias}},
  \bibinfo{author}{\bibfnamefont{X.}~\bibnamefont{Liu}}, \bibnamefont{and}
  \bibinfo{author}{\bibfnamefont{M.}~\bibnamefont{Nielsen}},
  \bibinfo{journal}{Phys. Rev. D} \textbf{\bibinfo{volume}{88}},
  \bibinfo{pages}{096014} (\bibinfo{year}{2013}), \eprint{1307.7100}.

\bibitem[{\citenamefont{Wang et~al.}(2021{\natexlab{a}})\citenamefont{Wang,
  Meng, and Zhu}}]{Wang:2020htx}
\bibinfo{author}{\bibfnamefont{B.}~\bibnamefont{Wang}},
  \bibinfo{author}{\bibfnamefont{L.}~\bibnamefont{Meng}}, \bibnamefont{and}
  \bibinfo{author}{\bibfnamefont{S.-L.} \bibnamefont{Zhu}},
  \bibinfo{journal}{Phys. Rev. D} \textbf{\bibinfo{volume}{103}},
  \bibinfo{pages}{L021501} (\bibinfo{year}{2021}{\natexlab{a}}),
  \eprint{2011.10922}.

\bibitem[{\citenamefont{Ozdem and Yildirim}(2021)}]{Ozdem:2021hka}
\bibinfo{author}{\bibfnamefont{U.}~\bibnamefont{Ozdem}} \bibnamefont{and}
  \bibinfo{author}{\bibfnamefont{A.~K.} \bibnamefont{Yildirim}}
  (\bibinfo{year}{2021}), \eprint{2104.13074}.

\bibitem[{\citenamefont{\"Ozdem and Azizi}(2021)}]{Ozdem:2021yvo}
\bibinfo{author}{\bibfnamefont{U.}~\bibnamefont{\"Ozdem}} \bibnamefont{and}
  \bibinfo{author}{\bibfnamefont{K.}~\bibnamefont{Azizi}}
  (\bibinfo{year}{2021}), \eprint{2102.09231}.

\bibitem[{\citenamefont{Wang}(2021{\natexlab{a}})}]{Wang:2020iqt}
\bibinfo{author}{\bibfnamefont{Z.-G.} \bibnamefont{Wang}},
  \bibinfo{journal}{Chin. Phys. C} \textbf{\bibinfo{volume}{45}},
  \bibinfo{pages}{073107} (\bibinfo{year}{2021}{\natexlab{a}}),
  \eprint{2011.10959}.

\bibitem[{\citenamefont{Azizi and Er}(2021)}]{Azizi:2020zyq}
\bibinfo{author}{\bibfnamefont{K.}~\bibnamefont{Azizi}} \bibnamefont{and}
  \bibinfo{author}{\bibfnamefont{N.}~\bibnamefont{Er}}, \bibinfo{journal}{Eur.
  Phys. J. C} \textbf{\bibinfo{volume}{81}}, \bibinfo{pages}{61}
  (\bibinfo{year}{2021}), \eprint{2011.11488}.

\bibitem[{\citenamefont{Wang}(2021{\natexlab{b}})}]{Wang:2020dgr}
\bibinfo{author}{\bibfnamefont{Z.-G.} \bibnamefont{Wang}},
  \bibinfo{journal}{Int. J. Mod. Phys. A} \textbf{\bibinfo{volume}{35}},
  \bibinfo{pages}{2150107} (\bibinfo{year}{2021}{\natexlab{b}}),
  \eprint{2012.11869}.

\bibitem[{\citenamefont{Xu et~al.}(2020)\citenamefont{Xu, Liu, Cui, and
  Huang}}]{Xu:2020evn}
\bibinfo{author}{\bibfnamefont{Y.-J.} \bibnamefont{Xu}},
  \bibinfo{author}{\bibfnamefont{Y.-L.} \bibnamefont{Liu}},
  \bibinfo{author}{\bibfnamefont{C.-Y.} \bibnamefont{Cui}}, \bibnamefont{and}
  \bibinfo{author}{\bibfnamefont{M.-Q.} \bibnamefont{Huang}}
  (\bibinfo{year}{2020}), \eprint{2011.14313}.

\bibitem[{\citenamefont{Wan and Qiao}(2021)}]{Wan:2020oxt}
\bibinfo{author}{\bibfnamefont{B.-D.} \bibnamefont{Wan}} \bibnamefont{and}
  \bibinfo{author}{\bibfnamefont{C.-F.} \bibnamefont{Qiao}},
  \bibinfo{journal}{Nucl. Phys. B} \textbf{\bibinfo{volume}{968}},
  \bibinfo{pages}{115450} (\bibinfo{year}{2021}), \eprint{2011.08747}.

\bibitem[{\citenamefont{Voloshin}(2019)}]{Voloshin:2019ilw}
\bibinfo{author}{\bibfnamefont{M.~B.} \bibnamefont{Voloshin}},
  \bibinfo{journal}{Phys. Lett. B} \textbf{\bibinfo{volume}{798}},
  \bibinfo{pages}{135022} (\bibinfo{year}{2019}), \eprint{1901.01936}.

\bibitem[{\citenamefont{Ferretti and Santopinto}(2020)}]{Ferretti:2020ewe}
\bibinfo{author}{\bibfnamefont{J.}~\bibnamefont{Ferretti}} \bibnamefont{and}
  \bibinfo{author}{\bibfnamefont{E.}~\bibnamefont{Santopinto}},
  \bibinfo{journal}{JHEP} \textbf{\bibinfo{volume}{04}}, \bibinfo{pages}{119}
  (\bibinfo{year}{2020}), \eprint{2001.01067}.

\bibitem[{\citenamefont{Yang et~al.}(2021{\natexlab{a}})\citenamefont{Yang,
  Ping, and Segovia}}]{Yang:2021zhe}
\bibinfo{author}{\bibfnamefont{G.}~\bibnamefont{Yang}},
  \bibinfo{author}{\bibfnamefont{J.}~\bibnamefont{Ping}}, \bibnamefont{and}
  \bibinfo{author}{\bibfnamefont{J.}~\bibnamefont{Segovia}}
  (\bibinfo{year}{2021}{\natexlab{a}}), \eprint{2109.04311}.

\bibitem[{\citenamefont{Chen et~al.}(2021)\citenamefont{Chen, Tan, and
  Chen}}]{Chen:2021uou}
\bibinfo{author}{\bibfnamefont{X.}~\bibnamefont{Chen}},
  \bibinfo{author}{\bibfnamefont{Y.}~\bibnamefont{Tan}}, \bibnamefont{and}
  \bibinfo{author}{\bibfnamefont{Y.}~\bibnamefont{Chen}},
  \bibinfo{journal}{Phys. Rev. D} \textbf{\bibinfo{volume}{104}},
  \bibinfo{pages}{014017} (\bibinfo{year}{2021}), \eprint{2103.07347}.

\bibitem[{\citenamefont{Giron et~al.}(2021)\citenamefont{Giron, Lebed, and
  Martinez}}]{Giron:2021sla}
\bibinfo{author}{\bibfnamefont{J.~F.} \bibnamefont{Giron}},
  \bibinfo{author}{\bibfnamefont{R.~F.} \bibnamefont{Lebed}}, \bibnamefont{and}
  \bibinfo{author}{\bibfnamefont{S.~R.} \bibnamefont{Martinez}},
  \bibinfo{journal}{Phys. Rev. D} \textbf{\bibinfo{volume}{104}},
  \bibinfo{pages}{054001} (\bibinfo{year}{2021}), \eprint{2106.05883}.

\bibitem[{\citenamefont{Shi et~al.}(2021)\citenamefont{Shi, Huang, and
  Wang}}]{Shi:2021jyr}
\bibinfo{author}{\bibfnamefont{P.-P.} \bibnamefont{Shi}},
  \bibinfo{author}{\bibfnamefont{F.}~\bibnamefont{Huang}}, \bibnamefont{and}
  \bibinfo{author}{\bibfnamefont{W.-L.} \bibnamefont{Wang}},
  \bibinfo{journal}{Phys. Rev. D} \textbf{\bibinfo{volume}{103}},
  \bibinfo{pages}{094038} (\bibinfo{year}{2021}), \eprint{2105.02397}.

\bibitem[{\citenamefont{Jin et~al.}(2020)\citenamefont{Jin, Liu, Xue, Huang,
  and Ping}}]{Jin:2020yjn}
\bibinfo{author}{\bibfnamefont{X.}~\bibnamefont{Jin}},
  \bibinfo{author}{\bibfnamefont{X.}~\bibnamefont{Liu}},
  \bibinfo{author}{\bibfnamefont{Y.}~\bibnamefont{Xue}},
  \bibinfo{author}{\bibfnamefont{H.}~\bibnamefont{Huang}}, \bibnamefont{and}
  \bibinfo{author}{\bibfnamefont{J.}~\bibnamefont{Ping}}
  (\bibinfo{year}{2020}), \eprint{2011.12230}.

\bibitem[{\citenamefont{Ebert et~al.}(2008)\citenamefont{Ebert, Faustov, and
  Galkin}}]{Ebert:2008kb}
\bibinfo{author}{\bibfnamefont{D.}~\bibnamefont{Ebert}},
  \bibinfo{author}{\bibfnamefont{R.~N.} \bibnamefont{Faustov}},
  \bibnamefont{and} \bibinfo{author}{\bibfnamefont{V.~O.}
  \bibnamefont{Galkin}}, \bibinfo{journal}{Eur. Phys. J. C}
  \textbf{\bibinfo{volume}{58}}, \bibinfo{pages}{399} (\bibinfo{year}{2008}),
  \eprint{0808.3912}.

\bibitem[{\citenamefont{Wu and Chen}(2021)}]{Wu:2021ezz}
\bibinfo{author}{\bibfnamefont{Q.}~\bibnamefont{Wu}} \bibnamefont{and}
  \bibinfo{author}{\bibfnamefont{D.-Y.} \bibnamefont{Chen}}
  (\bibinfo{year}{2021}), \eprint{2108.06700}.

\bibitem[{\citenamefont{Chen et~al.}(2013)\citenamefont{Chen, Liu, and
  Matsuki}}]{Chen:2013wca}
\bibinfo{author}{\bibfnamefont{D.-Y.} \bibnamefont{Chen}},
  \bibinfo{author}{\bibfnamefont{X.}~\bibnamefont{Liu}}, \bibnamefont{and}
  \bibinfo{author}{\bibfnamefont{T.}~\bibnamefont{Matsuki}},
  \bibinfo{journal}{Phys. Rev. Lett.} \textbf{\bibinfo{volume}{110}},
  \bibinfo{pages}{232001} (\bibinfo{year}{2013}), \eprint{1303.6842}.

\bibitem[{\citenamefont{Karliner and Rosner}(2021)}]{Karliner:2021qok}
\bibinfo{author}{\bibfnamefont{M.}~\bibnamefont{Karliner}} \bibnamefont{and}
  \bibinfo{author}{\bibfnamefont{J.~L.} \bibnamefont{Rosner}},
  \bibinfo{journal}{Phys. Rev. D} \textbf{\bibinfo{volume}{104}},
  \bibinfo{pages}{034033} (\bibinfo{year}{2021}), \eprint{2107.04915}.

\bibitem[{\citenamefont{Guo and Oller}(2021)}]{Guo:2020vmu}
\bibinfo{author}{\bibfnamefont{Z.-H.} \bibnamefont{Guo}} \bibnamefont{and}
  \bibinfo{author}{\bibfnamefont{J.~A.} \bibnamefont{Oller}},
  \bibinfo{journal}{Phys. Rev. D} \textbf{\bibinfo{volume}{103}},
  \bibinfo{pages}{054021} (\bibinfo{year}{2021}), \eprint{2012.11904}.

\bibitem[{\citenamefont{Sun and Xiao}(2020)}]{Sun:2020hjw}
\bibinfo{author}{\bibfnamefont{Z.-F.} \bibnamefont{Sun}} \bibnamefont{and}
  \bibinfo{author}{\bibfnamefont{C.-W.} \bibnamefont{Xiao}}
  (\bibinfo{year}{2020}), \eprint{2011.09404}.

\bibitem[{\citenamefont{Chen and Huang}(2021)}]{Chen:2020yvq}
\bibinfo{author}{\bibfnamefont{R.}~\bibnamefont{Chen}} \bibnamefont{and}
  \bibinfo{author}{\bibfnamefont{Q.}~\bibnamefont{Huang}},
  \bibinfo{journal}{Phys. Rev. D} \textbf{\bibinfo{volume}{103}},
  \bibinfo{pages}{034008} (\bibinfo{year}{2021}), \eprint{2011.09156}.

\bibitem[{\citenamefont{Ding et~al.}(2021)\citenamefont{Ding, Jiang, Song, and
  He}}]{Ding:2021igr}
\bibinfo{author}{\bibfnamefont{Z.-M.} \bibnamefont{Ding}},
  \bibinfo{author}{\bibfnamefont{H.-Y.} \bibnamefont{Jiang}},
  \bibinfo{author}{\bibfnamefont{D.}~\bibnamefont{Song}}, \bibnamefont{and}
  \bibinfo{author}{\bibfnamefont{J.}~\bibnamefont{He}}, \bibinfo{journal}{Eur.
  Phys. J. C} \textbf{\bibinfo{volume}{81}}, \bibinfo{pages}{732}
  (\bibinfo{year}{2021}), \eprint{2107.00855}.

\bibitem[{\citenamefont{Yang et~al.}(2021{\natexlab{b}})\citenamefont{Yang,
  Cao, Guo, Nieves, and Valderrama}}]{Yang:2020nrt}
\bibinfo{author}{\bibfnamefont{Z.}~\bibnamefont{Yang}},
  \bibinfo{author}{\bibfnamefont{X.}~\bibnamefont{Cao}},
  \bibinfo{author}{\bibfnamefont{F.-K.} \bibnamefont{Guo}},
  \bibinfo{author}{\bibfnamefont{J.}~\bibnamefont{Nieves}}, \bibnamefont{and}
  \bibinfo{author}{\bibfnamefont{M.~P.} \bibnamefont{Valderrama}},
  \bibinfo{journal}{Phys. Rev. D} \textbf{\bibinfo{volume}{103}},
  \bibinfo{pages}{074029} (\bibinfo{year}{2021}{\natexlab{b}}),
  \eprint{2011.08725}.

\bibitem[{\citenamefont{Meng et~al.}(2020)\citenamefont{Meng, Wang, and
  Zhu}}]{Meng:2020ihj}
\bibinfo{author}{\bibfnamefont{L.}~\bibnamefont{Meng}},
  \bibinfo{author}{\bibfnamefont{B.}~\bibnamefont{Wang}}, \bibnamefont{and}
  \bibinfo{author}{\bibfnamefont{S.-L.} \bibnamefont{Zhu}},
  \bibinfo{journal}{Phys. Rev. D} \textbf{\bibinfo{volume}{102}},
  \bibinfo{pages}{111502} (\bibinfo{year}{2020}), \eprint{2011.08656}.

\bibitem[{\citenamefont{Ikeno et~al.}(2021)\citenamefont{Ikeno, Molina, and
  Oset}}]{Ikeno:2020mra}
\bibinfo{author}{\bibfnamefont{N.}~\bibnamefont{Ikeno}},
  \bibinfo{author}{\bibfnamefont{R.}~\bibnamefont{Molina}}, \bibnamefont{and}
  \bibinfo{author}{\bibfnamefont{E.}~\bibnamefont{Oset}},
  \bibinfo{journal}{Phys. Lett. B} \textbf{\bibinfo{volume}{814}},
  \bibinfo{pages}{136120} (\bibinfo{year}{2021}), \eprint{2011.13425}.

\bibitem[{\citenamefont{Dong et~al.}(2021{\natexlab{a}})\citenamefont{Dong,
  Guo, and Zou}}]{Dong:2020hxe}
\bibinfo{author}{\bibfnamefont{X.-K.} \bibnamefont{Dong}},
  \bibinfo{author}{\bibfnamefont{F.-K.} \bibnamefont{Guo}}, \bibnamefont{and}
  \bibinfo{author}{\bibfnamefont{B.-S.} \bibnamefont{Zou}},
  \bibinfo{journal}{Phys. Rev. Lett.} \textbf{\bibinfo{volume}{126}},
  \bibinfo{pages}{152001} (\bibinfo{year}{2021}{\natexlab{a}}),
  \eprint{2011.14517}.

\bibitem[{\citenamefont{Dong et~al.}(2021{\natexlab{b}})\citenamefont{Dong,
  Guo, and Zou}}]{Dong:2021bvy}
\bibinfo{author}{\bibfnamefont{X.-K.} \bibnamefont{Dong}},
  \bibinfo{author}{\bibfnamefont{F.-K.} \bibnamefont{Guo}}, \bibnamefont{and}
  \bibinfo{author}{\bibfnamefont{B.-S.} \bibnamefont{Zou}}
  (\bibinfo{year}{2021}{\natexlab{b}}), \eprint{2108.02673}.

\bibitem[{\citenamefont{Dong et~al.}(2021{\natexlab{c}})\citenamefont{Dong,
  Guo, and Zou}}]{Dong:2021juy}
\bibinfo{author}{\bibfnamefont{X.-K.} \bibnamefont{Dong}},
  \bibinfo{author}{\bibfnamefont{F.-K.} \bibnamefont{Guo}}, \bibnamefont{and}
  \bibinfo{author}{\bibfnamefont{B.-S.} \bibnamefont{Zou}},
  \bibinfo{journal}{Progr. Phys.} \textbf{\bibinfo{volume}{41}},
  \bibinfo{pages}{65} (\bibinfo{year}{2021}{\natexlab{c}}),
  \eprint{2101.01021}.

\bibitem[{\citenamefont{Du et~al.}(2020)\citenamefont{Du, Wang, and
  Zhao}}]{Du:2020vwb}
\bibinfo{author}{\bibfnamefont{M.-C.} \bibnamefont{Du}},
  \bibinfo{author}{\bibfnamefont{Q.}~\bibnamefont{Wang}}, \bibnamefont{and}
  \bibinfo{author}{\bibfnamefont{Q.}~\bibnamefont{Zhao}}
  (\bibinfo{year}{2020}), \eprint{2011.09225}.

\bibitem[{\citenamefont{Yan et~al.}(2021)\citenamefont{Yan, Peng,
  S\'anchez~S\'anchez, and Pavon~Valderrama}}]{Yan:2021tcp}
\bibinfo{author}{\bibfnamefont{M.-J.} \bibnamefont{Yan}},
  \bibinfo{author}{\bibfnamefont{F.-Z.} \bibnamefont{Peng}},
  \bibinfo{author}{\bibfnamefont{M.}~\bibnamefont{S\'anchez~S\'anchez}},
  \bibnamefont{and}
  \bibinfo{author}{\bibfnamefont{M.}~\bibnamefont{Pavon~Valderrama}}
  (\bibinfo{year}{2021}), \eprint{2102.13058}.

\bibitem[{\citenamefont{Simonov}(2021)}]{Simonov:2020ozp}
\bibinfo{author}{\bibfnamefont{Y.~A.} \bibnamefont{Simonov}},
  \bibinfo{journal}{JHEP} \textbf{\bibinfo{volume}{04}}, \bibinfo{pages}{051}
  (\bibinfo{year}{2021}), \eprint{2011.12326}.

\bibitem[{\citenamefont{Zhu}(2021)}]{Zhu:2021vtd}
\bibinfo{author}{\bibfnamefont{K.}~\bibnamefont{Zhu}}, \bibinfo{journal}{Int.
  J. Mod. Phys. A} \textbf{\bibinfo{volume}{36}}, \bibinfo{pages}{2150126}
  (\bibinfo{year}{2021}), \eprint{2101.10622}.

\bibitem[{\citenamefont{Nieves and Valderrama}(2012)}]{Nieves:2012tt}
\bibinfo{author}{\bibfnamefont{J.}~\bibnamefont{Nieves}} \bibnamefont{and}
  \bibinfo{author}{\bibfnamefont{M.~P.} \bibnamefont{Valderrama}},
  \bibinfo{journal}{Phys. Rev. D} \textbf{\bibinfo{volume}{86}},
  \bibinfo{pages}{056004} (\bibinfo{year}{2012}), \eprint{1204.2790}.

\bibitem[{\citenamefont{Hidalgo-Duque et~al.}(2013)\citenamefont{Hidalgo-Duque,
  Nieves, and Valderrama}}]{HidalgoDuque:2012pq}
\bibinfo{author}{\bibfnamefont{C.}~\bibnamefont{Hidalgo-Duque}},
  \bibinfo{author}{\bibfnamefont{J.}~\bibnamefont{Nieves}}, \bibnamefont{and}
  \bibinfo{author}{\bibfnamefont{M.~P.} \bibnamefont{Valderrama}},
  \bibinfo{journal}{Phys. Rev. D} \textbf{\bibinfo{volume}{87}},
  \bibinfo{pages}{076006} (\bibinfo{year}{2013}), \eprint{1210.5431}.

\bibitem[{\citenamefont{Guo et~al.}(2013)\citenamefont{Guo, Hidalgo-Duque,
  Nieves, and Valderrama}}]{Guo:2013sya}
\bibinfo{author}{\bibfnamefont{F.-K.} \bibnamefont{Guo}},
  \bibinfo{author}{\bibfnamefont{C.}~\bibnamefont{Hidalgo-Duque}},
  \bibinfo{author}{\bibfnamefont{J.}~\bibnamefont{Nieves}}, \bibnamefont{and}
  \bibinfo{author}{\bibfnamefont{M.~P.} \bibnamefont{Valderrama}},
  \bibinfo{journal}{Phys. Rev. D} \textbf{\bibinfo{volume}{88}},
  \bibinfo{pages}{054007} (\bibinfo{year}{2013}), \eprint{1303.6608}.

\bibitem[{\citenamefont{Guo et~al.}(2018)\citenamefont{Guo, Hanhart,
  Mei\ss{}ner, Wang, Zhao, and Zou}}]{Guo:2017jvc}
\bibinfo{author}{\bibfnamefont{F.-K.} \bibnamefont{Guo}},
  \bibinfo{author}{\bibfnamefont{C.}~\bibnamefont{Hanhart}},
  \bibinfo{author}{\bibfnamefont{U.-G.} \bibnamefont{Mei\ss{}ner}},
  \bibinfo{author}{\bibfnamefont{Q.}~\bibnamefont{Wang}},
  \bibinfo{author}{\bibfnamefont{Q.}~\bibnamefont{Zhao}}, \bibnamefont{and}
  \bibinfo{author}{\bibfnamefont{B.-S.} \bibnamefont{Zou}},
  \bibinfo{journal}{Rev. Mod. Phys.} \textbf{\bibinfo{volume}{90}},
  \bibinfo{pages}{015004} (\bibinfo{year}{2018}), \eprint{1705.00141}.

\bibitem[{\citenamefont{Yuan et~al.}(2008)}]{Belle:2007dwu}
\bibinfo{author}{\bibfnamefont{C.~Z.} \bibnamefont{Yuan}} \bibnamefont{et~al.}
  (\bibinfo{collaboration}{Belle}), \bibinfo{journal}{Phys. Rev. D}
  \textbf{\bibinfo{volume}{77}}, \bibinfo{pages}{011105}
  (\bibinfo{year}{2008}), \eprint{0709.2565}.

\bibitem[{\citenamefont{Shen et~al.}(2014)}]{Belle:2014fgf}
\bibinfo{author}{\bibfnamefont{C.~P.} \bibnamefont{Shen}} \bibnamefont{et~al.}
  (\bibinfo{collaboration}{Belle}), \bibinfo{journal}{Phys. Rev. D}
  \textbf{\bibinfo{volume}{89}}, \bibinfo{pages}{072015}
  (\bibinfo{year}{2014}), \eprint{1402.6578}.

\bibitem[{\citenamefont{Ablikim et~al.}(2018)}]{BESIII:2018iop}
\bibinfo{author}{\bibfnamefont{M.}~\bibnamefont{Ablikim}} \bibnamefont{et~al.}
  (\bibinfo{collaboration}{BESIII}), \bibinfo{journal}{Phys. Rev. D}
  \textbf{\bibinfo{volume}{97}}, \bibinfo{pages}{071101}
  (\bibinfo{year}{2018}), \eprint{1802.01216}.

\bibitem[{\citenamefont{Aaij et~al.}(2021{\natexlab{b}})}]{LHCb:2021uow}
\bibinfo{author}{\bibfnamefont{R.}~\bibnamefont{Aaij}} \bibnamefont{et~al.}
  (\bibinfo{collaboration}{LHCb}), \bibinfo{journal}{Phys. Rev. Lett.}
  \textbf{\bibinfo{volume}{127}}, \bibinfo{pages}{082001}
  (\bibinfo{year}{2021}{\natexlab{b}}), \eprint{2103.01803}.

\bibitem[{\citenamefont{Ortega et~al.}(2021)\citenamefont{Ortega, Entem, and
  Fernandez}}]{Ortega:2021enc}
\bibinfo{author}{\bibfnamefont{P.~G.} \bibnamefont{Ortega}},
  \bibinfo{author}{\bibfnamefont{D.~R.} \bibnamefont{Entem}}, \bibnamefont{and}
  \bibinfo{author}{\bibfnamefont{F.}~\bibnamefont{Fernandez}},
  \bibinfo{journal}{Phys. Lett. B} \textbf{\bibinfo{volume}{818}},
  \bibinfo{pages}{136382} (\bibinfo{year}{2021}), \eprint{2103.07871}.

\bibitem[{\citenamefont{Meng et~al.}(2021)\citenamefont{Meng, Wang, Wang, and
  Zhu}}]{Meng:2021rdg}
\bibinfo{author}{\bibfnamefont{L.}~\bibnamefont{Meng}},
  \bibinfo{author}{\bibfnamefont{B.}~\bibnamefont{Wang}},
  \bibinfo{author}{\bibfnamefont{G.-J.} \bibnamefont{Wang}}, \bibnamefont{and}
  \bibinfo{author}{\bibfnamefont{S.-L.} \bibnamefont{Zhu}}
  (\bibinfo{year}{2021}), \eprint{2104.08469}.

\bibitem[{\citenamefont{Aaij et~al.}(2018)}]{LHCb:2018oeg}
\bibinfo{author}{\bibfnamefont{R.}~\bibnamefont{Aaij}} \bibnamefont{et~al.}
  (\bibinfo{collaboration}{LHCb}), \bibinfo{journal}{Eur. Phys. J. C}
  \textbf{\bibinfo{volume}{78}}, \bibinfo{pages}{1019} (\bibinfo{year}{2018}),
  \eprint{1809.07416}.

\bibitem[{\citenamefont{Chilikin et~al.}(2014)}]{Belle:2014nuw}
\bibinfo{author}{\bibfnamefont{K.}~\bibnamefont{Chilikin}} \bibnamefont{et~al.}
  (\bibinfo{collaboration}{Belle}), \bibinfo{journal}{Phys. Rev. D}
  \textbf{\bibinfo{volume}{90}}, \bibinfo{pages}{112009}
  (\bibinfo{year}{2014}), \eprint{1408.6457}.

\bibitem[{\citenamefont{Cao and Dai}(2019{\natexlab{a}})}]{Cao:2018vmv}
\bibinfo{author}{\bibfnamefont{X.}~\bibnamefont{Cao}} \bibnamefont{and}
  \bibinfo{author}{\bibfnamefont{J.-P.} \bibnamefont{Dai}},
  \bibinfo{journal}{Phys. Rev. D} \textbf{\bibinfo{volume}{100}},
  \bibinfo{pages}{054004} (\bibinfo{year}{2019}{\natexlab{a}}),
  \eprint{1811.06434}.

\bibitem[{\citenamefont{Wang et~al.}(2021{\natexlab{b}})\citenamefont{Wang,
  Zhou, Liu, and Matsuki}}]{Wang:2020kej}
\bibinfo{author}{\bibfnamefont{J.-Z.} \bibnamefont{Wang}},
  \bibinfo{author}{\bibfnamefont{Q.-S.} \bibnamefont{Zhou}},
  \bibinfo{author}{\bibfnamefont{X.}~\bibnamefont{Liu}}, \bibnamefont{and}
  \bibinfo{author}{\bibfnamefont{T.}~\bibnamefont{Matsuki}},
  \bibinfo{journal}{Eur. Phys. J. C} \textbf{\bibinfo{volume}{81}},
  \bibinfo{pages}{51} (\bibinfo{year}{2021}{\natexlab{b}}),
  \eprint{2011.08628}.

\bibitem[{\citenamefont{Guo et~al.}(2020)\citenamefont{Guo, Liu, and
  Sakai}}]{Guo:2019twa}
\bibinfo{author}{\bibfnamefont{F.-K.} \bibnamefont{Guo}},
  \bibinfo{author}{\bibfnamefont{X.-H.} \bibnamefont{Liu}}, \bibnamefont{and}
  \bibinfo{author}{\bibfnamefont{S.}~\bibnamefont{Sakai}},
  \bibinfo{journal}{Prog. Part. Nucl. Phys.} \textbf{\bibinfo{volume}{112}},
  \bibinfo{pages}{103757} (\bibinfo{year}{2020}), \eprint{1912.07030}.

\bibitem[{\citenamefont{Ge et~al.}(2021)\citenamefont{Ge, Liu, and
  Ke}}]{Ge:2021sdq}
\bibinfo{author}{\bibfnamefont{Y.-H.} \bibnamefont{Ge}},
  \bibinfo{author}{\bibfnamefont{X.-H.} \bibnamefont{Liu}}, \bibnamefont{and}
  \bibinfo{author}{\bibfnamefont{H.-W.} \bibnamefont{Ke}}
  (\bibinfo{year}{2021}), \eprint{2103.05282}.

\bibitem[{\citenamefont{Cao et~al.}(2021)\citenamefont{Cao, Dai, and
  Yang}}]{Cao:2020cfx}
\bibinfo{author}{\bibfnamefont{X.}~\bibnamefont{Cao}},
  \bibinfo{author}{\bibfnamefont{J.-P.} \bibnamefont{Dai}}, \bibnamefont{and}
  \bibinfo{author}{\bibfnamefont{Z.}~\bibnamefont{Yang}},
  \bibinfo{journal}{Eur. Phys. J. C} \textbf{\bibinfo{volume}{81}},
  \bibinfo{pages}{184} (\bibinfo{year}{2021}), \eprint{2011.09244}.

\bibitem[{\citenamefont{Yang and Guo}(2021)}]{Yang:2021jof}
\bibinfo{author}{\bibfnamefont{Z.}~\bibnamefont{Yang}} \bibnamefont{and}
  \bibinfo{author}{\bibfnamefont{F.-K.} \bibnamefont{Guo}}
  (\bibinfo{year}{2021}), \eprint{2107.12247}.

\bibitem[{\citenamefont{Liu et~al.}(2021)\citenamefont{Liu, Chen, and
  He}}]{Liu:2021ojf}
\bibinfo{author}{\bibfnamefont{J.}~\bibnamefont{Liu}},
  \bibinfo{author}{\bibfnamefont{D.-Y.} \bibnamefont{Chen}}, \bibnamefont{and}
  \bibinfo{author}{\bibfnamefont{J.}~\bibnamefont{He}} (\bibinfo{year}{2021}),
  \eprint{2108.00148}.

\bibitem[{\citenamefont{Liu and Oka}(2016)}]{Liu:2016dli}
\bibinfo{author}{\bibfnamefont{X.-H.} \bibnamefont{Liu}} \bibnamefont{and}
  \bibinfo{author}{\bibfnamefont{M.}~\bibnamefont{Oka}},
  \bibinfo{journal}{Nucl. Phys. A} \textbf{\bibinfo{volume}{954}},
  \bibinfo{pages}{352} (\bibinfo{year}{2016}), \eprint{1602.07069}.

\bibitem[{\citenamefont{Cao and Dai}(2019{\natexlab{b}})}]{Cao:2019kst}
\bibinfo{author}{\bibfnamefont{X.}~\bibnamefont{Cao}} \bibnamefont{and}
  \bibinfo{author}{\bibfnamefont{J.-p.} \bibnamefont{Dai}},
  \bibinfo{journal}{Phys. Rev. D} \textbf{\bibinfo{volume}{100}},
  \bibinfo{pages}{054033} (\bibinfo{year}{2019}{\natexlab{b}}),
  \eprint{1904.06015}.

\bibitem[{\citenamefont{Albaladejo et~al.}(2016)\citenamefont{Albaladejo, Guo,
  Hidalgo-Duque, and Nieves}}]{Albaladejo:2015lob}
\bibinfo{author}{\bibfnamefont{M.}~\bibnamefont{Albaladejo}},
  \bibinfo{author}{\bibfnamefont{F.-K.} \bibnamefont{Guo}},
  \bibinfo{author}{\bibfnamefont{C.}~\bibnamefont{Hidalgo-Duque}},
  \bibnamefont{and} \bibinfo{author}{\bibfnamefont{J.}~\bibnamefont{Nieves}},
  \bibinfo{journal}{Phys. Lett. B} \textbf{\bibinfo{volume}{755}},
  \bibinfo{pages}{337} (\bibinfo{year}{2016}), \eprint{1512.03638}.

\bibitem[{\citenamefont{Aaij et~al.}(2013)}]{LHCb:2013kpp}
\bibinfo{author}{\bibfnamefont{R.}~\bibnamefont{Aaij}} \bibnamefont{et~al.}
  (\bibinfo{collaboration}{LHCb}), \bibinfo{journal}{Phys. Rev. D}
  \textbf{\bibinfo{volume}{87}}, \bibinfo{pages}{072004}
  (\bibinfo{year}{2013}), \eprint{1302.1213}.

\bibitem[{\citenamefont{Aaij et~al.}(2015)}]{LHCb:2015yax}
\bibinfo{author}{\bibfnamefont{R.}~\bibnamefont{Aaij}} \bibnamefont{et~al.}
  (\bibinfo{collaboration}{LHCb}), \bibinfo{journal}{Phys. Rev. Lett.}
  \textbf{\bibinfo{volume}{115}}, \bibinfo{pages}{072001}
  (\bibinfo{year}{2015}), \eprint{1507.03414}.

\bibitem[{\citenamefont{Aaij et~al.}(2019{\natexlab{a}})}]{LHCb:2019kea}
\bibinfo{author}{\bibfnamefont{R.}~\bibnamefont{Aaij}} \bibnamefont{et~al.}
  (\bibinfo{collaboration}{LHCb}), \bibinfo{journal}{Phys. Rev. Lett.}
  \textbf{\bibinfo{volume}{122}}, \bibinfo{pages}{222001}
  (\bibinfo{year}{2019}{\natexlab{a}}), \eprint{1904.03947}.

\bibitem[{\citenamefont{Cheng and Chua}(2015)}]{Cheng:2015cca}
\bibinfo{author}{\bibfnamefont{H.-Y.} \bibnamefont{Cheng}} \bibnamefont{and}
  \bibinfo{author}{\bibfnamefont{C.-K.} \bibnamefont{Chua}},
  \bibinfo{journal}{Phys. Rev. D} \textbf{\bibinfo{volume}{92}},
  \bibinfo{pages}{096009} (\bibinfo{year}{2015}), \eprint{1509.03708}.

\bibitem[{\citenamefont{Aaij et~al.}(2019{\natexlab{b}})}]{LHCb:2019maw}
\bibinfo{author}{\bibfnamefont{R.}~\bibnamefont{Aaij}} \bibnamefont{et~al.}
  (\bibinfo{collaboration}{LHCb}), \bibinfo{journal}{Phys. Rev. Lett.}
  \textbf{\bibinfo{volume}{122}}, \bibinfo{pages}{152002}
  (\bibinfo{year}{2019}{\natexlab{b}}), \eprint{1901.05745}.

\bibitem[{\citenamefont{Aaij et~al.}(2017{\natexlab{a}})}]{LHCb:2016axx}
\bibinfo{author}{\bibfnamefont{R.}~\bibnamefont{Aaij}} \bibnamefont{et~al.}
  (\bibinfo{collaboration}{LHCb}), \bibinfo{journal}{Phys. Rev. Lett.}
  \textbf{\bibinfo{volume}{118}}, \bibinfo{pages}{022003}
  (\bibinfo{year}{2017}{\natexlab{a}}), \eprint{1606.07895}.

\bibitem[{\citenamefont{Aaij et~al.}(2017{\natexlab{b}})}]{LHCb:2016nsl}
\bibinfo{author}{\bibfnamefont{R.}~\bibnamefont{Aaij}} \bibnamefont{et~al.}
  (\bibinfo{collaboration}{LHCb}), \bibinfo{journal}{Phys. Rev. D}
  \textbf{\bibinfo{volume}{95}}, \bibinfo{pages}{012002}
  (\bibinfo{year}{2017}{\natexlab{b}}), \eprint{1606.07898}.

\bibitem[{\citenamefont{Aaltonen et~al.}(2009)}]{CDF:2009jgo}
\bibinfo{author}{\bibfnamefont{T.}~\bibnamefont{Aaltonen}} \bibnamefont{et~al.}
  (\bibinfo{collaboration}{CDF}), \bibinfo{journal}{Phys. Rev. Lett.}
  \textbf{\bibinfo{volume}{102}}, \bibinfo{pages}{242002}
  (\bibinfo{year}{2009}), \eprint{0903.2229}.

\bibitem[{\citenamefont{Aaltonen et~al.}(2017)}]{CDF:2011pep}
\bibinfo{author}{\bibfnamefont{T.}~\bibnamefont{Aaltonen}} \bibnamefont{et~al.}
  (\bibinfo{collaboration}{CDF}), \bibinfo{journal}{Mod. Phys. Lett. A}
  \textbf{\bibinfo{volume}{32}}, \bibinfo{pages}{1750139}
  (\bibinfo{year}{2017}), \eprint{1101.6058}.

\bibitem[{\citenamefont{Chatrchyan et~al.}(2014)}]{CMS:2013jru}
\bibinfo{author}{\bibfnamefont{S.}~\bibnamefont{Chatrchyan}}
  \bibnamefont{et~al.} (\bibinfo{collaboration}{CMS}), \bibinfo{journal}{Phys.
  Lett. B} \textbf{\bibinfo{volume}{734}}, \bibinfo{pages}{261}
  (\bibinfo{year}{2014}), \eprint{1309.6920}.

\bibitem[{\citenamefont{Yang et~al.}(2017)\citenamefont{Yang, Wang, and
  Mei\ss{}ner}}]{Yang:2017nde}
\bibinfo{author}{\bibfnamefont{Z.}~\bibnamefont{Yang}},
  \bibinfo{author}{\bibfnamefont{Q.}~\bibnamefont{Wang}}, \bibnamefont{and}
  \bibinfo{author}{\bibfnamefont{U.-G.} \bibnamefont{Mei\ss{}ner}},
  \bibinfo{journal}{Phys. Lett. B} \textbf{\bibinfo{volume}{775}},
  \bibinfo{pages}{50} (\bibinfo{year}{2017}), \eprint{1706.00960}.

\bibitem[{\citenamefont{Zyla et~al.}(2020)}]{Zyla:2020zbs}
\bibinfo{author}{\bibfnamefont{P.~A.} \bibnamefont{Zyla}} \bibnamefont{et~al.}
  (\bibinfo{collaboration}{Particle Data Group}), \bibinfo{journal}{PTEP}
  \textbf{\bibinfo{volume}{2020}}, \bibinfo{pages}{083C01}
  (\bibinfo{year}{2020}).

\bibitem[{\citenamefont{Guo et~al.}(2011)\citenamefont{Guo, Hanhart, Li,
  Meissner, and Zhao}}]{Guo:2010ak}
\bibinfo{author}{\bibfnamefont{F.-K.} \bibnamefont{Guo}},
  \bibinfo{author}{\bibfnamefont{C.}~\bibnamefont{Hanhart}},
  \bibinfo{author}{\bibfnamefont{G.}~\bibnamefont{Li}},
  \bibinfo{author}{\bibfnamefont{U.-G.} \bibnamefont{Meissner}},
  \bibnamefont{and} \bibinfo{author}{\bibfnamefont{Q.}~\bibnamefont{Zhao}},
  \bibinfo{journal}{Phys. Rev. D} \textbf{\bibinfo{volume}{83}},
  \bibinfo{pages}{034013} (\bibinfo{year}{2011}), \eprint{1008.3632}.

\bibitem[{\citenamefont{Guo}(2020)}]{Guo:2020oqk}
\bibinfo{author}{\bibfnamefont{F.-K.} \bibnamefont{Guo}},
  \bibinfo{journal}{Nucl. Phys. Rev.} \textbf{\bibinfo{volume}{37}},
  \bibinfo{pages}{406} (\bibinfo{year}{2020}), \eprint{2001.05884}.

\bibitem[{\citenamefont{Altmannshofer et~al.}(2019)}]{Belle-II:2018jsg}
\bibinfo{author}{\bibfnamefont{W.}~\bibnamefont{Altmannshofer}}
  \bibnamefont{et~al.} (\bibinfo{collaboration}{Belle-II}),
  \bibinfo{journal}{PTEP} \textbf{\bibinfo{volume}{2019}},
  \bibinfo{pages}{123C01} (\bibinfo{year}{2019}), \bibinfo{note}{[Erratum: PTEP
  2020, 029201 (2020)]}, \eprint{1808.10567}.

\bibitem[{\citenamefont{Bondar et~al.}(2012)}]{Belle:2011aa}
\bibinfo{author}{\bibfnamefont{A.}~\bibnamefont{Bondar}} \bibnamefont{et~al.}
  (\bibinfo{collaboration}{Belle}), \bibinfo{journal}{Phys. Rev. Lett.}
  \textbf{\bibinfo{volume}{108}}, \bibinfo{pages}{122001}
  (\bibinfo{year}{2012}), \eprint{1110.2251}.

\bibitem[{\citenamefont{Aaij et~al.}(2020{\natexlab{a}})}]{Aaij:2020hon}
\bibinfo{author}{\bibfnamefont{R.}~\bibnamefont{Aaij}} \bibnamefont{et~al.}
  (\bibinfo{collaboration}{LHCb}), \bibinfo{journal}{Phys. Rev. Lett.}
  \textbf{\bibinfo{volume}{125}}, \bibinfo{pages}{242001}
  (\bibinfo{year}{2020}{\natexlab{a}}), \eprint{2009.00025}.

\bibitem[{\citenamefont{Aaij et~al.}(2020{\natexlab{b}})}]{Aaij:2020ypa}
\bibinfo{author}{\bibfnamefont{R.}~\bibnamefont{Aaij}} \bibnamefont{et~al.}
  (\bibinfo{collaboration}{LHCb}), \bibinfo{journal}{Phys. Rev. D}
  \textbf{\bibinfo{volume}{102}}, \bibinfo{pages}{112003}
  (\bibinfo{year}{2020}{\natexlab{b}}), \eprint{2009.00026}.

\bibitem[{\citenamefont{Baru et~al.}(2021)\citenamefont{Baru, Epelbaum, Filin,
  Hanhart, and Nefediev}}]{Baru:2021ddn}
\bibinfo{author}{\bibfnamefont{V.}~\bibnamefont{Baru}},
  \bibinfo{author}{\bibfnamefont{E.}~\bibnamefont{Epelbaum}},
  \bibinfo{author}{\bibfnamefont{A.~A.} \bibnamefont{Filin}},
  \bibinfo{author}{\bibfnamefont{C.}~\bibnamefont{Hanhart}}, \bibnamefont{and}
  \bibinfo{author}{\bibfnamefont{A.~V.} \bibnamefont{Nefediev}}
  (\bibinfo{year}{2021}), \eprint{2110.00398}.

\end{thebibliography}

\end{document}